# GEARBOXES : INDIRECT IDENTIFICATION OF DYNAMIC FORCES TRANSMITTED TO HOUSING THROUGH BEARINGS

Gu Xiang ZHANG *, Emmanuel RIGAUD **, Jean-Claude PASCAL ***, Jean SABOT **

\* Département Acoustique Industrielle, CETIM. 52, Avenue Félix Louat.
  BP 67. 60304 SENLIS Cedex, France

\*\* Laboratoire de Tribologie et Dynamique des Systèmes UMR 5513.
  Ecole Centrale de Lyon. BP 163. 69131 ECULLY Cedex, France

\*\*\* ENSIM, Université du Maine, BP 535. 72017 LE MANS, France

## INTRODUCTION

The noise radiated by a gearbox is due to the vibrations of its housing. These vibrations come from an internal excitation generated by the meshing process (static transmission error [1]). Under operating conditions, this internal excitation generates a dynamic mesh force which is then transmitted to the housing through the shafts and bearings.

The prediction of the dynamic behaviour of the gearbox housing needs an accurate identification of the static transmission error or an identification of generalized forces acting on the housing. However, it is difficult to obtain these forces directly by experimental or numerical method. The objective of this paper is to develop an indirect method permitting to identify the dynamic forces transmitted to the housing through bearings. For this purpose, both numerical simulations and experimental investigations will be done. The numerical method allows to compute the generalized forces transmitted to the housing and to determine which forces are responsible for the vibrations of the housing. The experimental one enables us to identify the excitation forces applied on the housing from the measurement of transfer functions and vibration response of the housing. The comparison between the results obtained by the two approaches shows the effectiveness of the proposed indirect force identification methodology.

## DESCRIPTION OF THE GEARBOX

A single stage gearbox fitted out with a helical gear pair has been built. The main characteristics of the gear pair are similar to those of a car gearbox (table 1). Each shaft is mounted on two tapered roller bearings. The steel housing (650x420x150 mm) is of the right-angled parallelepiped form made up of five 40 mm thick faces (frame) and one thin face (300x450 mm) of 6 mm thickness. Fixed to the frame, this thin plate holds up two bearings supporting respectively the input and the output shafts.

By a flexible coupling, a motor is connected to the input shaft which allows to generate an input speed from 0 up to 3000 rpm and a torque from 0 up to 240 Nm. Power is absorbed by a generator connected to the output shaft via flexible couplings and belts. The space in front of the 6 mm thick face is free.

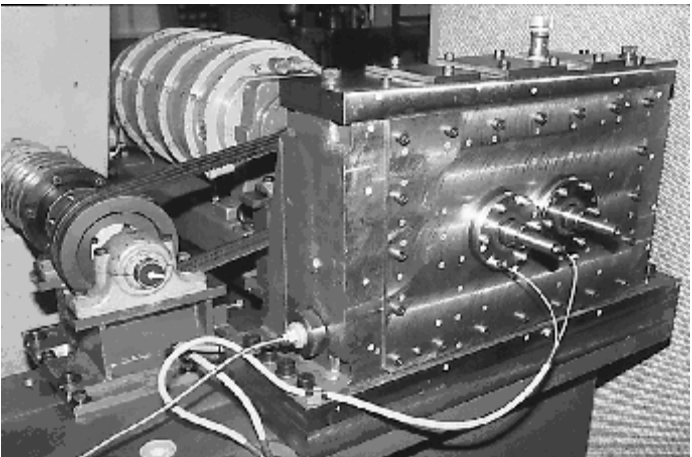

| | |
|---|---:|
| Number of teeth | 17 / 71 |
| Base radii (mm) | 23.4 / 97.7 |
| Normal module (mm) | 2.676 |
| Transverse pressure angle | 22° |
| Base helix angle | 24° |
| Transverse contact ratio | 1.718 |
| Overlap contact ratio | 1.030 |
| Total contact ratio | 2.748 |
| Facewidth (mm) | 20 |
| Centre distance (mm) | 128 |

**Figure 1** : Photo of the transmission system     **Table 1** : Characteristics of the gear pair.

**NUMERICAL APPROACH**

# I. MODELLING OF THE WHOLE GEARBOX AND COMPUTATION OF ITS DYNAMIC RESPONSE

The gearbox is modeled using a finite element method. The pinion and the gear are modeled with concentrated masses and rotary inertia. A 12x12 stiffness matrix is introduced to couple the 6 d.o.f. of the pinion to the 6 d.o.f. of the gear. The shafts are discretized using beam elements. The motor and load are connected to the shafts using rotary inertia and torsion stiffness elements. The housing is discretized using solid elements (40 mm thick faces) and shell elements (6 mm thick face). A 10x10 stiffness matrix is introduced to model tapered roller bearings with use of a method described in [2]. The bearings are discretized using solid elements. Totally, 1800 elements and 10000 d.o.f. are used for the whole gearbox.

As excitation source, only the transmission error is retained here, which can result from manufacturing errors, tooth modifications and from periodic mesh stiffness. Periodic static transmission error and mesh stiffness can be evaluated with help of a method described in [3].

For a constant mean speed of rotation, the equation which governs vibrations of the previously discretized gearbox can be written as follows :

$$\underline{\underline{M}}.\underline{\ddot{X}} + \underline{\underline{C}}.\underline{\dot{X}} + \underline{\underline{K}}.\underline{X} + k(t).\underline{\underline{D}}.\underline{X} = \underline{E}(t) \qquad (1)$$

M and K are the classical mass and stiffness matrix provided by the finite element method; the matrix D is derived from geometric characteristics of the gear pair; k(t) is the periodic mesh stiffness and E(t) is an equivalent force vector induced by the static transmission error. A modal base of the gearbox can be defined from the time-invariant homogeneous counterpart equation. Equation (1) is solved using a spectral and iterative method [4], which provides a direct spectral description of the vibration response at each d.o.f. and allows to evaluate simultaneously the dynamic mesh force, the generalized forces transmitted to the housing through the bearings and the vibration response of the housing. For all the numerical results proposed in this paper, viscous modal damping ratio is chosen equal to 1 % for all the natural modes.

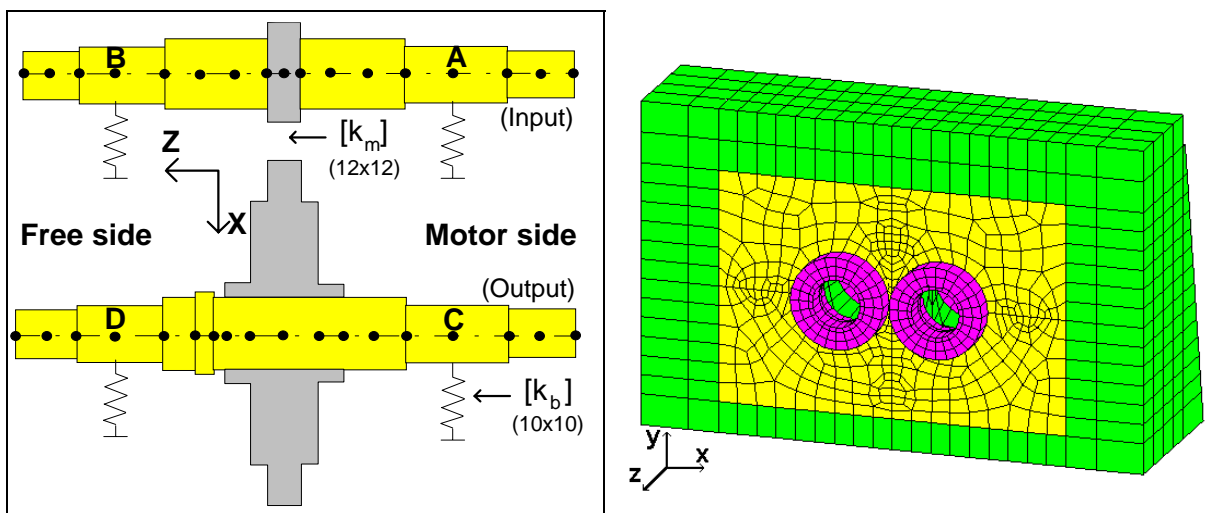

**Figure 2** : Modelling of the whole gearbox.

## II.   NUMERICAL RESULTS

### 2.1. Generalized forces transmitted to the bearings and vibration response of the housing

From the model of the whole gearbox and the spectral and iterative method, radial forces $F_x$ and $F_y$, axial force $F_z$, moments $M_x$ and $M_y$ transmitted to the housing through the bearings A, B, C and D and vibration response of the housing are calculated. Figures 3, 4 and 5 display the forces $F_z$ and moments $M_x$ and $M_y$ transmitted through bearings B and D against mesh frequency.

Figure 6 gives the calculated time- and space-averaged mean square vibration velocity of the housing $\langle \overline{v^2(\omega)} \rangle_S$ against mesh frequency.

$$\langle V^2(\omega) \rangle_S = \frac{1}{S} \int_S \left( \frac{1}{T} \int_T V^2(M,\omega,t)\,dt \right) dS \tag{2}$$

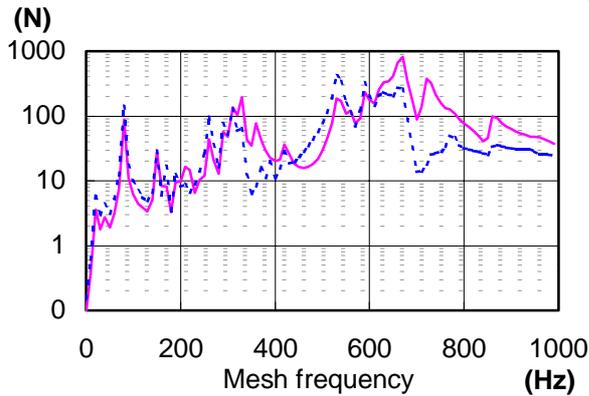

**Figure 3** : Axial forces $F_z(B)$ (———) and $F_z(D)$ (------) transmitted to the housing.

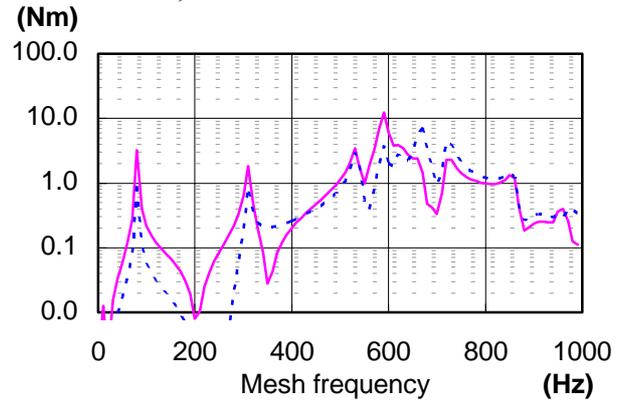

**Figure 4** : Moments $M_x(B)$ (———) and $M_x(D)$ (------) transmitted to the housing.

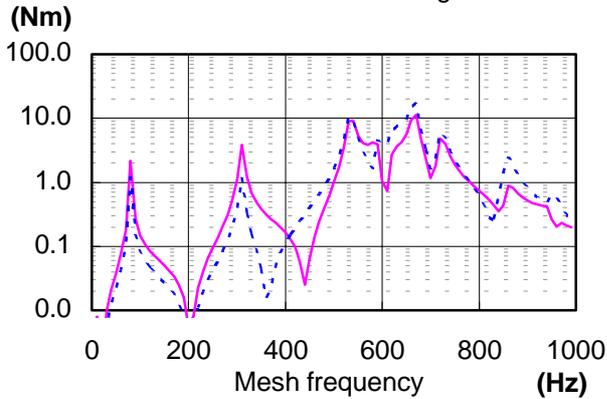

**Figure 5** : Moments $M_y(B)$ (———) and $M_y(D)$ (------) Transmitted to the housing.

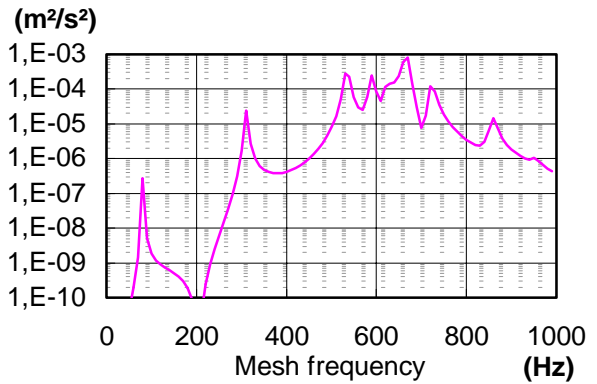

**Figure 6** : Time- and space-averaged mean square vibration velocity of housing $\langle \overline{v^2(\omega)} \rangle_S$.

### 2.2 Contribution of generalized forces to the vibration response of the housing

The vibration response of the housing can be rebuilt using the reciprocity principle : gear and shafts are suppressed and all the forces transmitted through the bearings are applied on the empty housing (in substitution of the static transmission error). Figure 9 shows that the vibration response of the housing calculated with this method is identical to that obtained directly by the spectral and iterative method in § 2.1.

It's also interesting to evaluate the relative contribution of the generalized forces transmitted through bearing A, B, C and D to the vibration response of the housing. Figure 7 shows that the vibration response of the housing is induced principally by the generalized forces transmitted through bearings B and D (bearings situated on the free side and supported by the 6 mm thick face). In fact, the quadratic mobility of the housing within 10-1000 Hz is much higher when the excitation force is applied on the 6 mm face than when the excitation force is applied on the 40 mm thick face (figure 8).

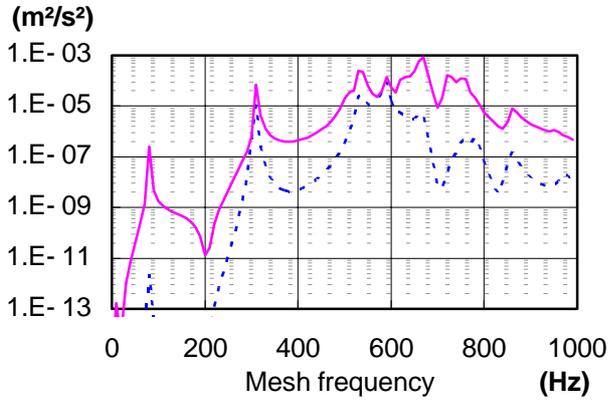
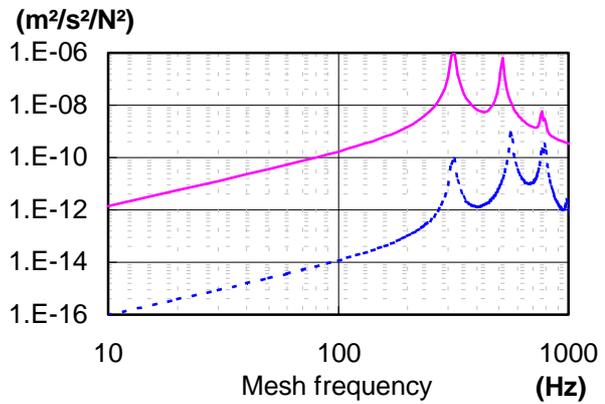

**Figure 7** : Housing response $\langle \overline{v^2(\omega)} \rangle_S$ induced by forces transmitted through bearings supported by 6 mm thick face (———) or by 40 mm thick face (------).

**Figure 8 :** Quadratic mobility of the housing. $F_z(B)=1N$ applied on 6 mm thick face (———). $F_z(A)=1N$ applied on 40 mm thick face (------).

The contribution of each force ($F_x$, $F_y$, $F_z$, $M_x$ and $M_y$) can also be evaluated. Figure 9 shows that the contribution of the radial forces $F_x$ and $F_y$ can be neglected. In fact, among the twenty generalized forces and moments transmitted to the housing through the bearings, the dominating components are firstly the axial forces $F_z(B)$ and $F_z(D)$ and secondarily the moments $M_x(B)$, $M_y(B)$, $M_x(D)$ and $M_y(D)$. The simulation on the time- and space-averaged mean square vibration velocity shows the vibration response of the housing induced by these six components is almost identical to that induced by the whole forces transmitted to the housing.

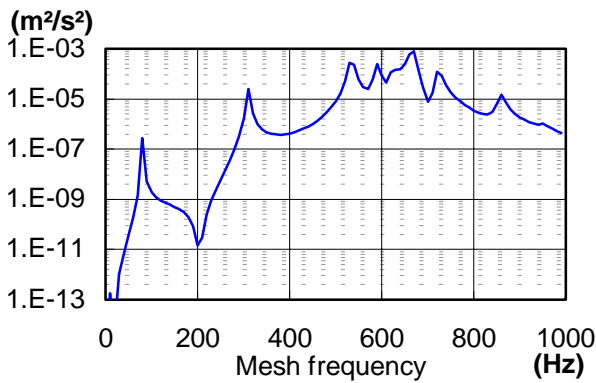
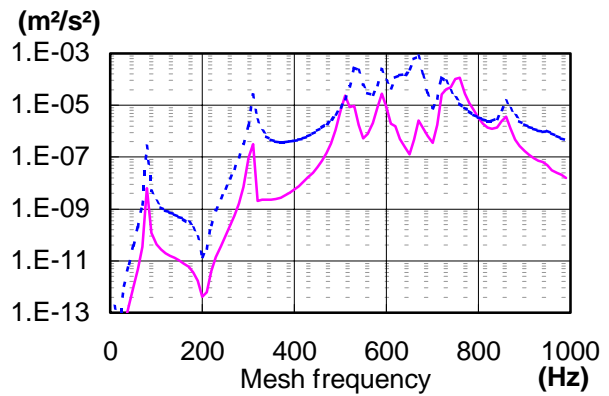

**Figure 9 :** Response $\langle \overline{v^2(\omega)} \rangle_S$ induced by the whole forces applied on the empty housing.

**Figure 10 :** Response $\langle \overline{v^2(\omega)} \rangle_S$ induced by the radial forces $F_x$ and $F_y$ (———).

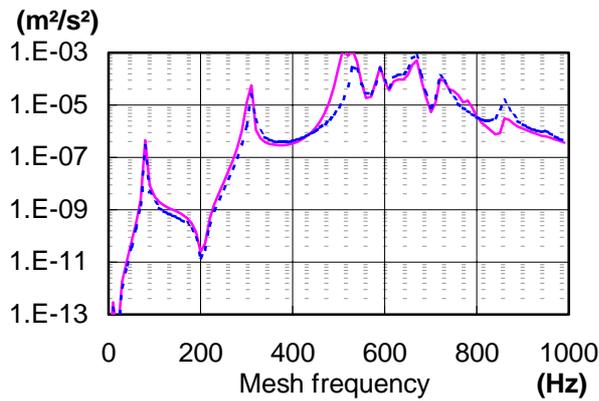 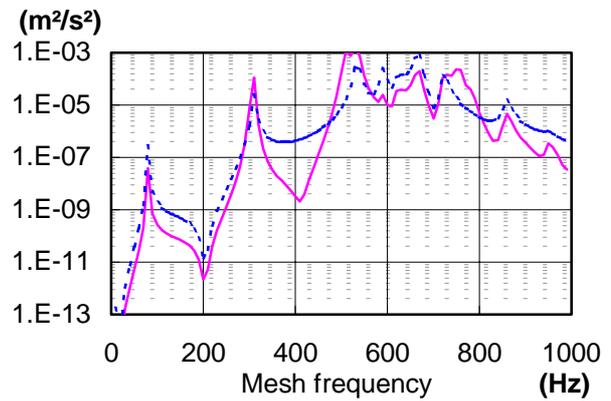

**Figure 11 :** Response $\langle \overline{v^2(\omega)} \rangle_S$ induced by the axial forces $F_z$ (———).

**Figure 12 :** Response $\langle \overline{v^2(\omega)} \rangle_S$ induced by the moments $M_x$ and $M_y$ (———).

## 2.3 Forces calculated with a rigid housing

Vibration response of housing is usually calculated via two uncoupled stages. First, the housing is supposed to be rigid and the generalized forces transmitted to the housing are evaluated by modelling of gear and shafts. Then, the calculated forces are applied on the empty elastic housing [5, 6] for computing the vibration response of the housing. In fact, mechanical properties of the housing modify both the natural frequencies of the gearbox and the transfer functions between static transmission error and forces transmitted through the bearings. As shown in figure 13, the forces transmitted through the bearings calculated with a rigid housing are very different from those calculated with an elastic housing. With this traditional method, the vibration response of the housing is considerably overestimated (figure 14).

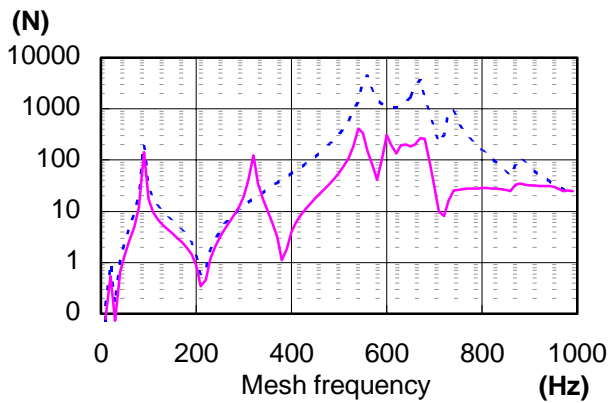 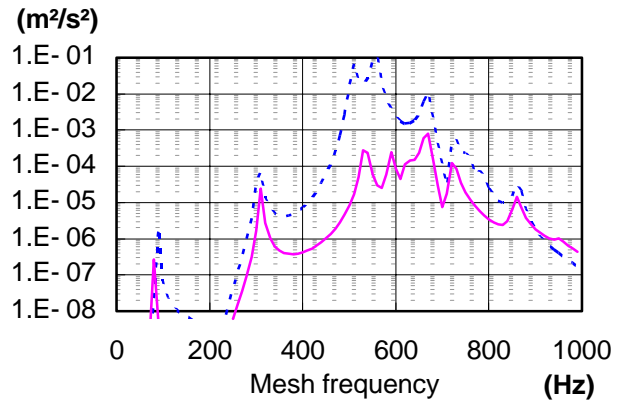

**Figure 13 :** Axial force $F_z(D)$ estimated with an elastic housing (———) or a rigid housing (-------).

**Figure 14 :** Response $\langle \overline{v^2(\omega)} \rangle_S$ induced by forces estimated with an elastic housing (———) or a rigid housing (--------).

In conclusion, the numerical simulation has confirmed that applying the generalized forces transmitted through the bearings on the empty housing allows to rebuild the housing response (reciprocity principle). Simulation results demonstrate the vibration response of the housing is induced mainly by the axial excitation forces, and secondarily by the moments $M_x$ and $M_y$ transmitted on the free side (the 6 mm thick face). Realistic estimation of these excitation forces needs the modelling of the whole gearbox taking into account the mechanical properties of the housing.

# EXPERIMENTAL APPROACH

It is demonstrated that the forces transmitted to the housing through the bearings can be evaluated by numerical methods. However, in practice, these forces can't be measured directly due to the impossibility for introducing force transducers. In the following, an inverse approach, which is based upon the vibration response measurement of the structure and FRF matrix inversion, is proposed here to identify experimentally the excitation forces.

## I.  MESURING EQUIVALENT FORCES BY INVERSE METHOD

As we know, the sound radiation of power transmitting system is caused principally by the vibrations of the housing of gearbox. We are only interested in the forces acting on the front plate of the gearbox. For utilisation of the inverse approach, some hypothesis must be made.

Firstly, suppose that the vibrations of the front plate of gearbox is produced merely by the forces transmitted to this plate through the bearings. In fact, this prototype was designed and constructed under the guideline that the front and rear plates of gearbox are mounted on a rigid frame realised with some welded steel plates of 40 mm's thickness, while the front plate is 6 mm thick. Moreover, a vibration investigation showed the vibration level of the front plate is much greater than that of the frame (within the interested frequency band, the difference between the average quadratic velocity of the frame and that of the front plate is 20 dB).

Secondly, assume that the forces acting on the front plate can be represented by some equivalent forces, which are punctual and oriented in the axial direction. Indeed, the previous numerical computation has given us evidence that the contribution of radial forces to the plate vibrations is much smaller than that of the axial ones. For simplifying the measurement, the radial forces will be ignored here.

As showed in figure 15, suppose that two groups of points are the positions of excitation forces : one group of 3 points, distinguished by G1, G2 and G3, are situated on the bearing supporting the driving shaft, the other group of 3 points D1, D2 and D3 are situated on the bearing supporting the driven shaft.

Several inverse methods enable us to determine indirectly the excitation forces. Their principle can be found in [7,8,9,10]. Experimentally, three steps are necessary. The first step consists of determining the FRF between the excitation positions and some response points chosen on the reception structure, a FRF matrix is hereby established. In the second step, the excitation forces being applied, the vibration responses of the reception structure are measured, a matrix of vibration responses can be constructed. Finally, by inverting the FRF matrix, the applied forces can be calculated.

In this article, the inverse method to be utilised is that called the inverse method of cross-spectral matrix of vibration responses, whose principle can be expressed as follows :

$$[G_{FF}] = [H]^+ [G_{aa}] [H]^{+H} ,  \qquad (3)$$

where,

$[G_{FF}]$ : cross-spectral matrix of responses consisting of auto-spectra and cross-spectra of the forces,

$[G_{aa}]$ : cross-spectral matrix of responses consisting of auto-spectra and cross-spectra of the responses,

$[H]$ : FRF matrix of the MIMO system.

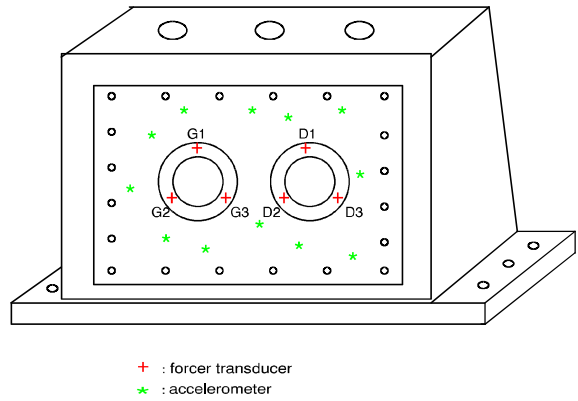

**Figure 15 :** Positions of the equivalent forces and selected vibration responses.

For the case reported here, 12 response points are randomly selected on the plate. First, while the gear transmission system is in silence, the FRF Acceleration/Force between the 6 equivalent forces and the 12 chosen vibration responses are characterised, and this FRF characterisation is realised by mean of a random excitation. Then, the machine is turned on, the vibration responses on the 12 points are measured simultaneously by 12 accelerometers. Both the auto-spectra of 12 vibration responses and the cross-spectra between them are acquired. According to equation (3), cross-spectral matrix of the equivalent forces can be finally obtained.

Two working cases were investigated. For one, the rotating speed is 500 rpm and the loading 80 Nm, for the other, the rotating speed is 1500 rpm and the loading 160 Nm. The auto-spectra of the 6 equivalent forces obtained by the inverse method for the two cases are given respectively in the figures 16 and 17. With this inverse method, the cross-spectra and the coherence coefficients between the equivalent forces are also calculable.

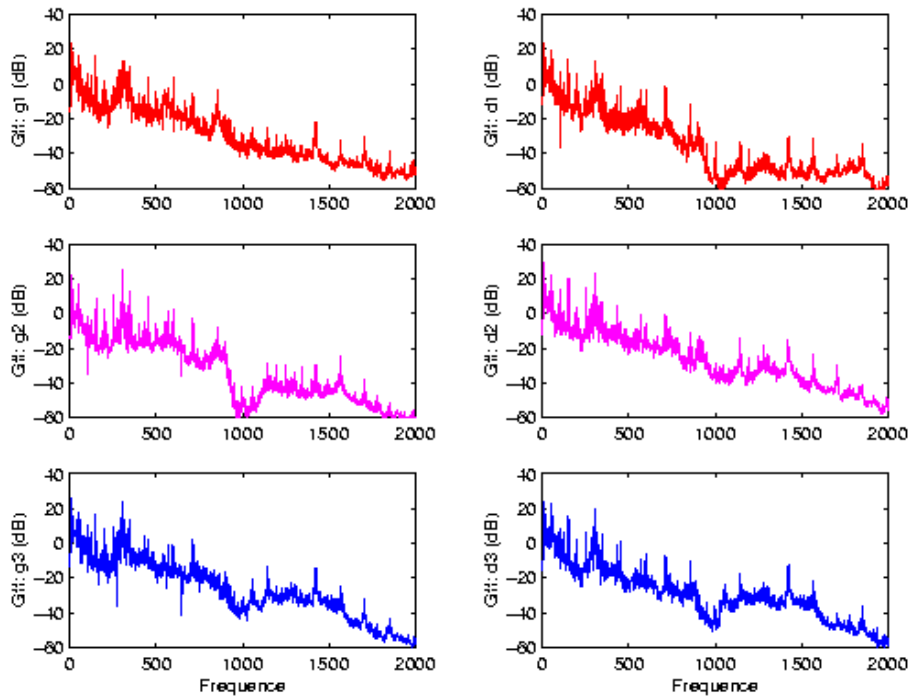

**Figure 16 :** Auto-spectra of the equivalent forces for the case 500 rpm, 80 Nm.

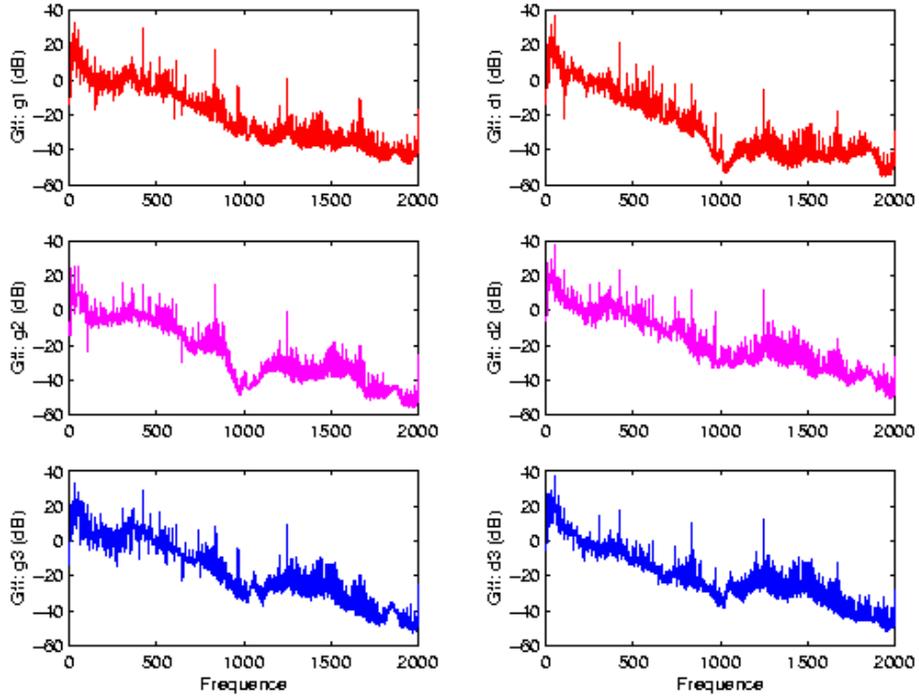

**Figure 17 :** Auto-spectra of the equivalent forces for the case 1500 rpm, 160 Nm.

## II    DETERMING RESULTING FORCES ET MOMENTS

In fact, the forces acting on the housing through bearings are some distributed forces introduced by internal mechanism. Beginning from the equivalent forces on the bearings calculated above, in the following, we'll represent the distributed force on each bearing by one resulting force and one moment.

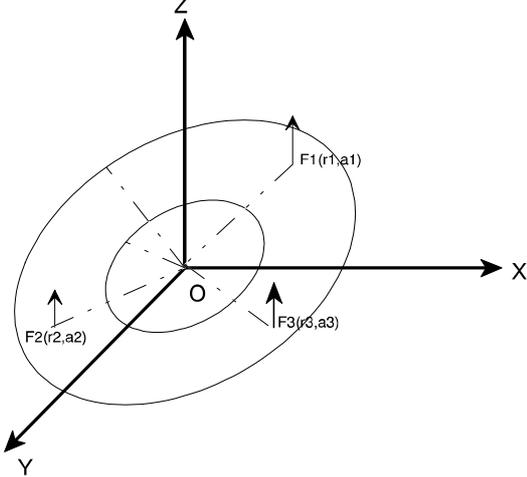

As illustrated in figure 4, on the surface of one bearing delimited by two radius R1 and R2, three equivalent punctual forces, oriented in the Z direction, are applied. If at a given frequency $\omega$ the three equivalent forces are perfectly coherent, one resulting force and one moment can be used to represent them. This resulting force, which is also oriented in the Z direction but applied at the original point O, can be deduced by summing directly the three forces.

**Figure 18 :** Resulting force and moment.

$$F_e = F_1 + F_2 + F_3 = |F_e|\, e^{j\phi}. \tag{4}$$

While, the resulting moment can be obtained as follows :

$$\begin{aligned}\vec{M}_e &= \vec{r}_1 \times F_1\, \vec{e}_z + \vec{r}_2 \times F_2\, \vec{e}_z + \vec{r}_3 \times F_3\, \vec{e}_z \\ &= -(F_1\, r_1 \cos\theta_1 + F_2 \times r_2 \cos\theta_2 + F_3 \times r_3 \cos\theta_3)\, \vec{e}_y + (F_1\, r_1 \sin\theta_1 + F_2\, r_2 \sin\theta_2 + F_3\, r_3 \sin\theta_3)\, \vec{e}_x \\ &= M_x\, \vec{e}_x + M_y\, \vec{e}_y\end{aligned} \tag{5}$$

So, the two components of the resulting moment can be expressed as :

$$M_x = F_1 r_1 \sin\theta_1 + F_2 r_2 \sin\theta_2 + F_3 r_3 \sin\theta_3 = |M_x| e^{j\phi_x}$$
$$M_y = -(F_1 r_1 \cos\theta_1 + F_2 r_2 \cos\theta_2 + F_3 r_3 \cos\theta_3) = |M_y| e^{j\phi_y}.$$
(6)

It should be pointed out here that the forces $F_1, F_2$ and $F_3$ are complex forces including both amplitude and phase. In case where the forces are totally coherent, these complex quantities can be calculated with help of a reference signal. If the auto-spectrum $G_{RR}(\omega)$ of reference and the cross-spectrum $G_{F_i R}(\omega)$ between the reference and the force $F_i$ are known, the complex force $F_i(\omega)$ can be calculated by :

$$F_i(\omega) = \frac{G_{F_i R}(\omega)}{\sqrt{G_{RR}(\omega)}}$$
(7)

In case where the three forces aren't totally coherent, a technique called Principal Component Analysis (PCA) must be employed to decompose the cross-spectral matrix of forces obtained by inverse method, a series of principal components can be thereby obtained [9]. For each principal component, it's necessary to determine its resulting force and moment. By summing the energy contributions of all the principal components, the auto-spectra of the total resulting force and moment can be evaluated.

It's found experimentally at the mesh frequency, the excitation forces are almost perfectly coherent. The table 2 presents the resulting forces and moments acting on the front plate at the mesh frequency 143 Hz for the case 500 rpm, 80 Nm, with comparison to the relevant results computed via the previous numerical method. And the table 3 gives the corresponding results at the mesh frequency 418 Hz for the case 1500 rpm and 160 Nm. For these two tables, the utilised reference signal is the force at the point $G_1$.

|  | Bearing supporting driving shaft | | Bearing supporting driven shaft | |
|---|---|---|---|---|
|  | Calculation | Experiment | Calculation | Experiment |
| $\vec{F}_e$ (Ampli. N) | 12 | 12 | 13 | 11 |
| $\vec{F}_e$ (Phase °) | 0° | 4° | -180° | -170° |
| $\vec{M}_x$ (Ampli. Nm) | 0.25 | 0.2 | 0.09 | 0.1 |
| $\vec{M}_x$ (Phase °) | 0° | -6° | 177° | 130° |
| $\vec{M}_y$ (Ampli. Nm) | 0.38 | 0.2 | 0.13 | 0.17 |
| $\vec{M}_y$ (Phase °) | 179° | 170° | 179° | 129° |

**Table 2 :** Amplitudes and phases of the resulting forces and moments at the mesh frequency $f_z$=143 Hz for the working case 500 rpm, 80 Nm.

|  | Bearing supporting driving shaft | | Bearing supporting driven shaft | |
|---|---|---|---|---|
|  | Calculation | Experiment | Calculation | Experiment |
| $\vec{F}_e$ (Ampli. N) | 37 | 52 | 28 | 19 |
| $\vec{F}_e$ (Phase °) | 65° | -14° | -163° | -165° |
| $\vec{M}_x$ (Ampli. Nm) | 0.55 | 0.8 | 0.55 | 0.35 |
| $\vec{M}_x$ (Phase °) | 19° | 18° | -158° | -148° |
| $\vec{M}_y$ (Ampli. Nm) | 0.2 | 1.2 | 0.4 | 0.7 |
| $\vec{M}_y$ (Phase °) | 31° | 143° | -162° | 100° |

**Table 3 :** Amplitudes and phases of the resulting forces and moments at the mesh frequency $f_z$=418 Hz for the working case 1500 rpm, 160 Nm.

Generally speaking, the agreement between the results obtained by numerical method and those by the experimental inverse method is good. The consideration of unique excitation source (transmission error) by the numerical simulation can explain in part the existing difference.

**CONCLUSION**

In order to determine indirectly the generalized forces applied on the gearbox, two approaches are tested, one is a numerical method, the other is experimental. For the numerical method, the vibration response of the gearbox can be evaluated with use of a finite element analysis and a spectral and iterative method. It's confirmed that applying the generalized forces transmitted through the bearings on the empty housing allows to rebuild the housing response (reciprocity principle). The simulation on the time- and space-averaged mean square vibration velocity shows the vibration response of the housing is induced principally by the axial exciting forces, and secondarily by the moments $M_x$ and $M_y$. transmitted on the free side (the 6 mm thick face). Realistic estimation of these excitation forces needs a modelling of the whole gearbox taking into account the mechanical properties of the housing.

For the experimental approach, an inverse method based upon the FRF measurement and characterisation of gearbox vibration response is developed. Beginning from the equivalent forces obtained by this inverse method, a resulting force and a moment are derived for representing the distributed force acting on each bearing. The comparison between the results obtained by these two methods shows the effectiveness of this indirect force determination methodology.

# REDUCTEURS A ENGRENAGES: IDENTIFICATION DES EFFORTS DYNAMIQUES TRANSMIS AUX CARTERS PAR LES PALIERS

Gu Xiang ZHANG *, Emmanuel RIGAUD **, Jean-Claude PASCAL ***, Jean SABOT **

* Département Acoustique Industrielle, CETIM. 52, Avenue Félix Louat. BP 67. 60304 SENLIS Cedex, France
** Laboratoire de Tribologie et Dynamique des Systèmes UMR 5513. Ecole Centrale de Lyon. BP 163. 69131 ECULLY Cedex, France
*** ENSIM, Université du Maine, BP 535. 72017 LE MANS, France

## RESUME

Le bruit engendré par le fonctionnement d'une transmission par engrenages est principalement induit par les vibrations de son carter. Ces vibrations ont pour origine principale l'erreur statique de transmission sous charge qui se traduit par une fluctuation des efforts s'exerçant sur les dents en prise. Par le corps des engrenages, les arbres et les paliers, ces excitations sont transmises au carter et l'excitent en vibration. En raison de la complexité de la structure, en général, l'observation directe de ces excitations par capteurs de force est impossible.

Deux approches sont proposées dans cette communication pour évaluer ces efforts, l'une est basée sur un modèle par éléments finis et l'autre expérimentale.

Pour l'approche "calcul numérique", une modélisation intégrant l'ensemble des composants de la transmission (engrenages, lignes d'arbres, roulement, carter et environnement) permet d'évaluer le transfert entre l'effort dynamique de denture et les efforts dynamiques transmis au carter par les roulements. Il est donc possible de connaître à partir de l'erreur statique de transmission sous charge les efforts généralisés transmis au carter par les paliers. Par cette approche numérique on peut également calculer la réponse vibratoire de la transmission complète et évaluer la contribution de chaque effort généralisé.

L'approche expérimentale utilise des méthodes inverses qui permettent de déterminer sur chaque palier des efforts équivalents induits par le roulement de l'arbre. L'expérimentation des méthodes inverses consiste d'abord à déterminer, machine au repos, les fonctions de transfert entre les points d'excitation et des points de réponse vibratoire sur le carter. Dans un deuxième temps, la machine étant en fonctionnement, on relève les réponses vibratoires aux points de réponse choisis. Par inversion de la matrice des fonctions de transfert, les excitations recherchées peuvent enfin être calculées. A partir des efforts équivalents obtenus, un effort et un moment résultants peuvent être quantifiés sur chaque boîtier de roulement.

Les deux approches proposées sont appliquées sur un banc d'essais des transmissions construit dans le cadre des activités du Club Transsil. L'analyse des résultats produits par l'approche expérimentale et la prédiction théorique a permis de vérifier la validé de ces deux approches et de démontrer la faisabilité d'une méthodologie visant à déterminer indirectement les efforts appliqués sur les carters des transmissions par engrenages.